\documentstyle[12pt]{article}
\raggedright  \parindent=20pt

\begin{document}
\pagestyle{empty}
\begin{flushright} CERN-TH.6731/92\\
KSUCNR-018-92\\
(revised)\\ \end{flushright}
\vspace*{5mm}
\begin{center}
{\bf Chemical equilibration of $\pi$, $\rho$ and $\omega$ mesons\\
in ultra--relativistic nuclear collisions} \\[10mm]
David Seibert \\[3mm] Theory Division, CERN, CH--1211 Geneva 23,
Switzerland$^*$
\\ and \\ Physics Department, Kent State University, Kent, OH 44240
USA\\[10mm]
{\bf Abstract} \\
\end{center}
I estimate the degree of local chemical equilibration for $\pi$,
$\rho$ and $\omega$ mesons in ultra--relativistic nuclear collisions.
The $\pi$ and $\rho$ mesons remain near chemical equilibrium in all
cases, while the $\omega$ meson density is typically 30--50\% higher
than the equilibrium value for O+Ag collisions at $\sqrt{s}=20$ GeV.
If chemical reactions are turned off, the $\omega$ meson density is
much larger, 2.5--3 times its equilibrium value.  Thus, $\omega$
mesons may provide the most sensitive tests for the degree of chemical
equilibration in nuclear collisions.\\
\vfill
\begin{flushleft} CERN-TH.6731/92 \\
KSUCNR-018-92 \\
November 1992 \\
(revised February 1993) \\ \end{flushleft}
\thispagestyle{empty}\mbox{}
\footnoterule
{$^*$Address until October 12, 1993; internet:
seibert@surya11.cern.ch.}
\newpage\setcounter{page}{1} \pagestyle{plain}

There exist proposals to measure the QCD transition temperature
[\ref{rrho0}] and the lifetime of the hadronic gas [\ref{ruli}] in
ultra--relativistic nuclear collisions, using dileptons from the
$\rho^0-\omega$ peak.  To determine the effectiveness of these and
similar proposals, one must know how close the behavior of the rapidly
expanding resonance gas is to that of a static system at fixed
temperature and chemical potential.  More specifically, one needs to
know how close the $\pi$, $\rho$ and $\omega$ distributions are to
their values at chemical and thermal equilibrium.

In this Letter, I estimate the degree of local chemical equilibration
that occurs in a central (low impact parameter) ultra--relativistic nuclear
collision.  This is  done by calculating chemical (number--changing)
reactions for an expanding resonance gas while assuming local thermal
equilibrium and entropy conservation, using several sets of possible
values for allowed four--body strong interaction reaction rates.  The
degree of equilibration obviously depends on the assumed reaction rates.
As these rates are not known, I use plausible values (around 1 fm$^2$--c
for exothermic reactions).  I neglect interactions with nucleons because
the meson density is much higher than the nucleon density in the central
region of the collision even at the lowest energies that I consider.
However, this approximation should be tested in a future paper, as the
cross sections for meson--nucleon interactions are typically somewhat
larger than those for meson--meson reactions.  I use the conventions
that $\hbar={\rm c}=k_B=1$.

I find that the $\pi$ and $\rho$ mesons are always near chemical
equilibrium.  For the highest energy (largest $\sqrt{s}$) collisions,
the $\omega$ mesons are also nearly equilibrated, but significant
departures from equilibrium  are predicted for collisions such as
O+Ag at $\sqrt{s}=20$ GeV.  Observation of these departures from
equilibrium would provide confirmation of the approach to chemical
equilibrium in ultra--relativistic nuclear collisions, and would
give a measure of chemical reaction rates in a hot resonance gas.

There are eight exothermic chemical reactions involving at most four
particles that are allowed in a system of $\pi$, $\rho$ and $\omega$
mesons.  Two are simple decays: $\rho \rightarrow \pi \pi$ and $\omega
\rightarrow \pi \pi \pi$, with rates $\Gamma_{\rho \rightarrow \pi \pi}
= 0.77$ fm$^{-1}$--c and $\Gamma_{\omega \rightarrow \pi \pi \pi} =
0.04$ fm$^{-1}$--c at $T=0$ [\ref{rPDG}].  I assume here that these
decay rates do not change with temperature; this may be a bad
assumption, but it simplifies the calculation, and the temperature
dependence is largely unknown.

The other six exothermic chemical reactions are $\omega \omega
\rightarrow \rho \rho$, $\omega \omega \rightarrow \pi \pi$, $\rho
\omega \rightarrow \pi \rho$, $\rho \rho \rightarrow \pi \omega$,
$\rho \rho \rightarrow \pi \pi$ and $\pi \omega \rightarrow \pi \pi$.
The rate constants for these reactions are all unknown, so I estimate
them in several different ways.  I obtain the rate
constants for the endothermic reverse reactions using the principle
of detailed balance (the equilibrium rates must be equal for the
reaction and the reverse reaction) [\ref{rSMech}]:
\begin{equation}
R_{r_1 \cdots r_l \rightarrow p_1 \cdots p_m} \prod_{i=1}^l
n^{\rm eq}_{r_i} ~=~ R_{p_1 \cdots p_m \rightarrow r_1 \cdots r_l}
\prod_{i=1}^m n^{\rm eq}_{p_i},
\end{equation}
where $n^{\rm eq}_x$ is the equilibrium density of species $x$.

I assume boost--invariant longitudinal expansion of the matter formed
in the collision, with no transverse expansion, following the simple
model of Bjorken [\ref{rBj}].  In the absence of viscosity and
discontinuities, the entropy of the system is constant.  I use this
to obtain the temperature of the system, $T$, as a function of proper
time, $\tau$, treating the system as an ideal gas of massive particles.
\begin{equation}
\frac {A \tau} {2 \pi^2 T} \sum_i g_i \int_0^{\infty} k^2 dk
\frac {\epsilon_i(k) \, + \, k^2/3\epsilon_i(k)}
{e^{\epsilon_i(k)/T} \, - \, 1} = {\rm const}. \label{eScon}
\end{equation}
Here $A$ is the cross--sectional area of the region of hot matter,
$g_i$ is the degeneracy of species $i$, and $\epsilon_i(k) =
\left( k^2+m_i^2 \right)^{1/2}$ where $m_i$ is the mass of species $i$.
I take $g_{\pi}=3$ (isospin), $g_{\rho}=9$ (isospin and polarization)
and $g_{\omega}=3$ (polarization), with $m_{\pi}=140$ MeV and
$m_{\rho}=m_{\omega}=775$ MeV.

In this paper, I make the optimistic assumption that all of the entropy
is generated at the beginning of the collision.  I then estimate the
final entropy using the facts that most of the final state particles
are pions, which are nearly massless, and that a massless ideal gas has
approximately 3.6 units of entropy per particle.  Thus,
\begin{equation}
\frac {A \tau} {2 \pi^2 T} \sum_i g_i \int_0^{\infty} k^2 dk
\frac {\epsilon_i(k) \, + \, k^2/3\epsilon_i(k)}
{e^{\epsilon_i(k)/T} \, - \, 1} = 3.6 \, dN/dy, \label{eScon2}
\end{equation}
where $dN/dy$ is the rapidity density of particles in the final state
(both charged and neutral).  This probably results in an overestimate of
the temperature, but it is  sufficient for the purposes of this letter.

Once $A$ and $dN/dy$ are specified, the only parameter needed to describe
the expansion of the matter is the hadronic transition temperature.  I
assume that, above some temperature $T_c \simeq 150-200$ MeV, the matter
is in either a deconfined or a chirally symmetric phase.  The transition
from this phase is modelled as occurring in thermal and chemical
equilibrium; this provides a reasonable starting point, as equilibration
rates are expected to be significantly faster in these more symmetric
phases than in normal hadronic matter.  This starting point is also
supported by data from pp collisions at $\sqrt{s}=27.5$ GeV
[\ref{rinit}], which is fit well by an equilibrium distribution at
$T=150-155$ MeV.

Once $T_c$ is specified, the initial proper time (immediately after the
transition to a resonance gas), $\tau_0$, is obtained from
eq.~(\ref{eScon2}).  At $\tau_0$, all densities have thermal and
chemical equilibrium values:
\begin{equation}
n^{\rm eq}_i(T_c) = \frac {g_i} {2 \pi^2} \int_0^{\infty}
\frac {k^2 dk} {e^{\epsilon_i(k)/T_c} \, - \, 1}. \label{eneq}
\end{equation}
The subsequent evolution of the system is described by the following set
of coupled equations:
\begin{equation}
\frac {dn_i} {d\tau} = - \left( \frac {1} {\tau} + \Gamma_{i \rightarrow X}
\right) n_i + \sum_j \Gamma_{j  \rightarrow iX} n_j - \sum_j R_{ij
\rightarrow X} n_i n_j + \sum_{j,k}  R_{jk \rightarrow iX} n_j n_k.
\end{equation}
Here final states $X$ are summed over, and there are two obvious additional
terms due to the reaction $\pi \pi \pi \rightarrow \omega$. Exothermic
reactions typically have weaker temperature dependences than endothermic
reactions, so I estimate the exothermic reaction rates geometrically and
derive the endothermic reactions using the principle of detailed balance,
as described earlier.  I obtain $T$ as a function of $\tau$ by solving
eq.~(\ref{eScon2}).  [In practice, it is most convenient to work in
reverse, calculating $\tau$ as a function of $T$.]

In Fig.~1, I show results with chemical reactions turned off ($R=0$)
for O+Ag collisions at $\sqrt{s}=20$ GeV and for Pb+Pb collisions at
$\sqrt{s}=200$ GeV, assuming that $T_c=200$ MeV.  For the lower energy
O+Ag collisions, the $\pi$ and $\rho$ meson densities are not very far
from their chemical equilibrium values, but the $\omega$ density quickly
rises to 2--3 times its equilibrium value.  In the higher energy Pb+Pb
collisions, even the $\omega$ meson density remains near its equilibrium
value.  The deviation from equilibrium is smaller when the chemical
reactions are turned on, so I do not bother to show results for the
higher energy collisions with reactions turned on.

I allow the gas to evolve for 80 fm/c for the O+Ag collisions and 500
fm/c for the Pb+Pb collisions; however, transverse expansion develops
with a time scale of about 10 fm/c, causing the densities to decrease
rapidly so that the interactions freeze out quickly after about 10
fm/c.  This long expansion is reasonable for models without transverse
expansion, as the final temperature is 100 MeV, and the system reaches
a temperature of 120 MeV halfway through the evolution.  If I allowed
transverse expansion, the system would not evolve as long; however,
the final volume would be the same (because of entropy conservation)
and the expansion time would be proportionately shorter, so the meson
densities would be approximately the same.

The simplest assumption is that all exothermic rate constants are
approximately the same.  I thus take
\begin{equation}
R_{ij \rightarrow kl} ~=~ R \label{eR1}
\end{equation}
for all exothermic reactions.  I show results in Fig.~2 for O+Ag
collisions at $\sqrt{s}=20$ GeV with $T_c=200$ MeV, taking $R=1.5$
fm$^2$--c ($\sigma \approx 15$ mb).  The $\omega$ mesons are still
noticeably further from equilibrium than the $\pi$ or $\rho$ mesons.
This is probably because the $\omega$ density  is low, so they are
not easily destroyed in two--body processes, and  their decay rate
is also small.  This effect does not change much for $T_c=150$ MeV.

One obvious improvement on this first assumption is to assume that the
total reaction rates for all pairs of particles are approximately the
same, but that the relative probabilities of the different exothermic
final states are given by the probability that the initial state has
the same quantum numbers (apart from energy and momentum) as the final
state.  For example, for the reaction $\omega \omega \rightarrow X$,
the possible final states are $\omega \omega$, $\rho \rho$ and $\pi
\pi$.  The relative probabilities of these final states are 1:1:1/9
respectively, as $\omega \omega$ and $\rho \rho$ can always be produced
from $\omega \omega$, while for $\pi \pi$ the vector polarizations must
combine to produce a scalar final state.  The exothermic reaction rates
are then
\begin{eqnarray}
R_{\omega \omega \rightarrow \rho \rho} ~=~ 9R^{\rm tot}/19, \\
R_{\omega \omega \rightarrow \pi \pi} ~=~ R^{\rm tot}/19, \\
R_{\rho \omega \rightarrow \pi \rho} ~=~ R^{\rm tot}/4, \\
R_{\rho \rho \rightarrow \pi \omega} ~=~ R^{\rm tot}/12, \\
R_{\rho \rho \rightarrow \pi \pi} ~=~ R^{\rm tot}/12, \\
R_{\pi \omega \rightarrow \pi \pi} ~=~ \left[1-(n^{\rm eq}_{\rho} /
4 n^{\rm eq}_{\pi}) \right] R^{\rm tot}/4.
\end{eqnarray}
$R_{\pi \omega \rightarrow \pi \pi}$ is temperature dependent because one
of the possible final states is $\rho \rho$, and this reaction is
endothermic.

In order to put an upper bound on the size of the expected effect, I
use $R^{\rm tot}=6$ fm$^2$--c ($\sigma \approx 60$ mb -- approximately
equal to the pp cross--section [\ref{rPDG}]).  I show the results in
Fig.~3 for O+Ag collisions at $\sqrt{s}=20$ GeV with $T_c=200$ and 150
MeV.  The $\omega$ mesons are even further from equilibrium than those
shown in Fig.~2, but this is not surprising as the reaction rates (and
thus the equilibration rates) are smaller.

This provides a signal for the existence of a hot, nearly equilibrated
resonance gas -- the size of the final state enhancement of $\omega$
mesons with respect to $\rho^0$ mesons (as they are almost degenerate
and thus have almost the same density in thermal and chemical
equilibrium).  This signal is large enough to be possibly measurable in
collisions of small nuclei at high (but not too high) energies, and is
probably significantly smaller than the value for a non--interacting
system (with $R=0$).  It may be difficult to observe, however, as
measurement depends on being able to separate final--state mesons from
those that decay during the earlier stages of the collision.  The most
likely way to see them is to reconstruct the mesons from the
final--state pions, but I will not attempt to consider the feasibility
of this procedure as it depends on details of the detectors used for
the measurement.

One possible criticism of this signal is that I have neglected changes
in the $\omega$ width at finite temperature.  However, this is certainly
not true, as the chemical reaction terms I use are equivalent to
collisional broadening.  The magnitude of the collisional broadening
that I use here is a bit larger than the calculated change in the width
[\ref{rshuryak}], so the results of this simulation suggest that it
should be possible to see the change in the $\omega$ width
As the $\omega$ excess exists throughout most of the collision, it may be
possible to measure the integrated $\omega$ and $\rho$ densities rather
than the final--state densities.  This could be accomplished with the
dilepton signals, if the two peaks could be separated.  However, this is
probably at least as difficult as detecting the resonances from analysis
of final--state pions, so I again leave the feasibility of this technique
as an open problem for the experimenters.

The reported enhancement of $\phi$ mesons in nuclear collisions [\ref{rphi}]
may be related to the departure of $\omega$ mesons from equilibrium that
is seen here, as shown by Heinz and Lee [\ref{ruli}].  The $\phi$ lifetime
(50 fm/c) is even longer than that of the $\omega$, so it will be further
from equilibrium if there are no chemical reactions.  However, I have not
investigated systems including $\phi$ mesons, so I do not have any quantitative
results at the present time.

Finally, while $\pi$ and $\rho$ mesons are likely to be nearly in chemical
equilibrium even in collisions of relatively light ions at relatively low
energies, this is not true for $\omega$ mesons.  The departure of the
$\omega$ mesons from chemical equilibrium is small for collisions of large
nuclei at high energies, but care will be necessary when comparing to current
results at relatively low energies.  However, even under the most pessimistic
conditions considered (O+Ag collisions at $\sqrt{s}=20$ GeV with $R=0$ and
$T_c=200$ MeV), the $\pi$ and $\rho$ meson densities were fairly close to
their equilibrium values (factors of roughly 1.3 and 1.6 higher
respectively).

\section*{ Acknowledgements }

I thank Dr.\ P.V. Ruuskanen for interesting me in meson physics, Dr.\ K.
Rummukainen for helping with the graphs, and Drs.\ M. Jacob and K. Kajantie
for their comments on the manuscript.  This work was partially supported by
the US Department of Energy under Grant No.\ DOE/DE-FG02-86ER-40251.
This material is partially  based upon work supported by the North Atlantic
Treaty Organization under a Grant awarded in 1991.

%\vfill \eject

\section*{ References }

\begin{enumerate}

\item D. Seibert, Phys.\ Rev.\ Lett.\ {\bf 68}, 1476 (1992); D. Seibert,
V.K. Mishra and G. Fai, Phys.\ Rev.\ C {\bf 46}, 330 (1992). \label{rrho0}

\item U. Heinz and K.--S. Lee, Phys.\ Lett.\ B {\bf 259}, 162 (1991).
\label{ruli}

\item Particle Data Group, Review of Particle Properties, Phys.\ Rev.\
D {\bf 45}, Part 2, 1 (1992). \label{rPDG}

\item See, e.g., F. Reif, {\it Fundamentals of Statistical and Thermal
Physics} (McGraw--Hill, New York, 1965). \label{rSMech}

\item J.D. Bjorken, Phys.\ Rev.\ D {\bf 27}, 140 (1983). \label{rBj}

\item LEBC--EHS Collaboration, M. Aguilar--Benitez {\it et al.}, Z. Phys.\
C {\bf 50}, 405 (1991). \label{rinit}

\item E. Shuryak, Nucl.\ Phys.\ {\bf A533}, 761 (1991). \label{rshuryak}

\item NA38 Collaboration: C. Baglin {\it et al.}, Phys.\ Lett.\ B {\bf 272},
449 (1991); R. Ferreira, Nucl.\ Phys.\ {\bf A544}, 497c (1992). \label{rphi}

\end{enumerate}

%\vfill \eject

\section*{ Figure captions }

\begin{enumerate}

\item $R=0$, $T_c=200$ MeV: (a) O+Ag at $\sqrt{s}=20$ GeV ($\tau_0=4.6$ fm/c),
(b) Pb+Pb at $\sqrt{s}=200$ GeV ($\tau_0=27.8$ fm/c).

\item $R=1.5$ fm$^2$--c, $T_c=200$ MeV: O+Ag at $\sqrt{s}=20$ GeV
($\tau_0=4.6$ fm/c).

\item $R^{\rm tot}=6$ fm$^2$--c: (a) O+Ag at $\sqrt{s}=20$ GeV for
$T_c=200$ MeV ($\tau_0=4.6$ fm/c), (b) O+Ag at $\sqrt{s}=20$ GeV for
$T_c=150$ MeV ($\tau_0=15.8$ fm/c).

\end{enumerate}

\vfill \eject

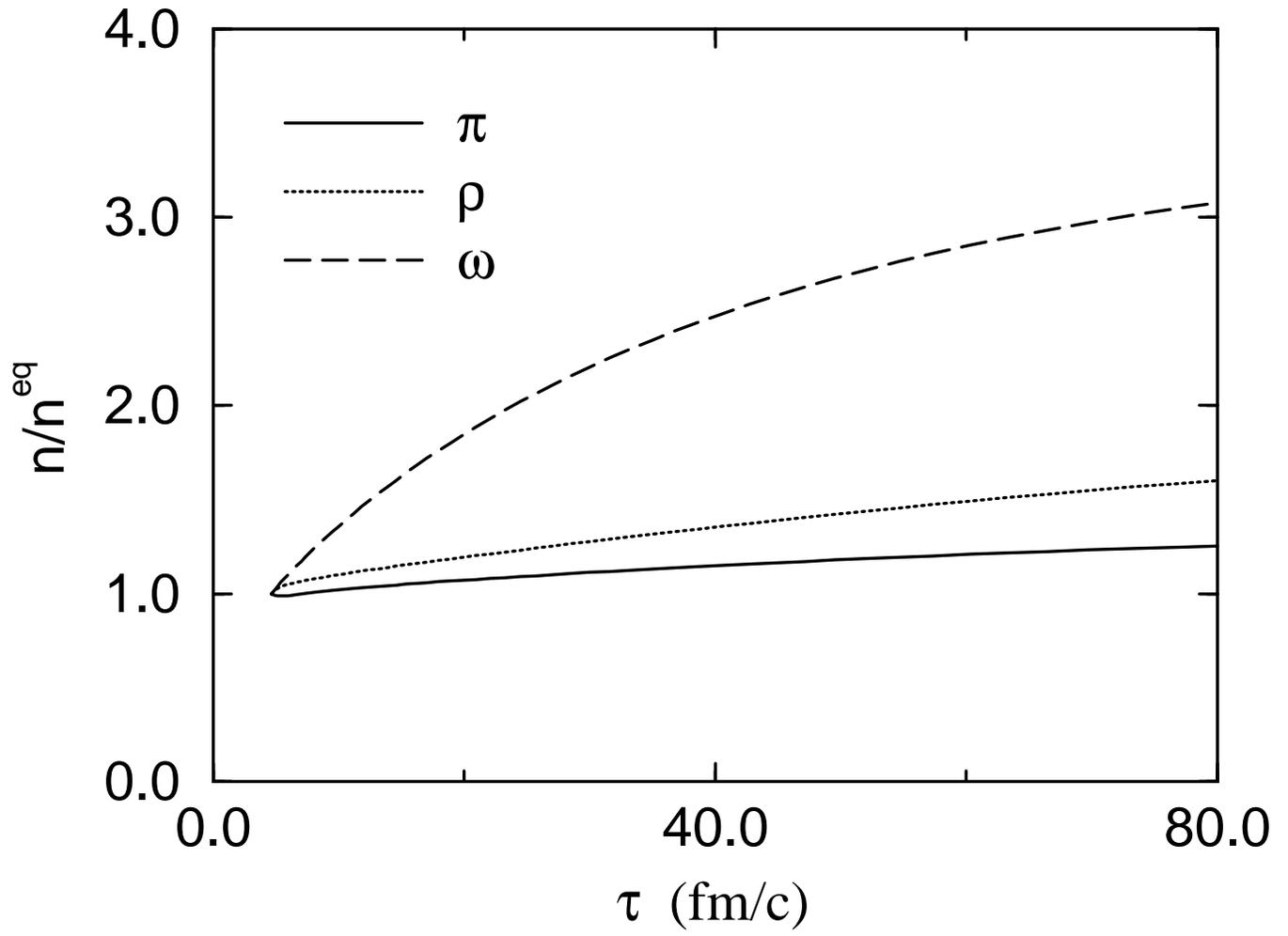
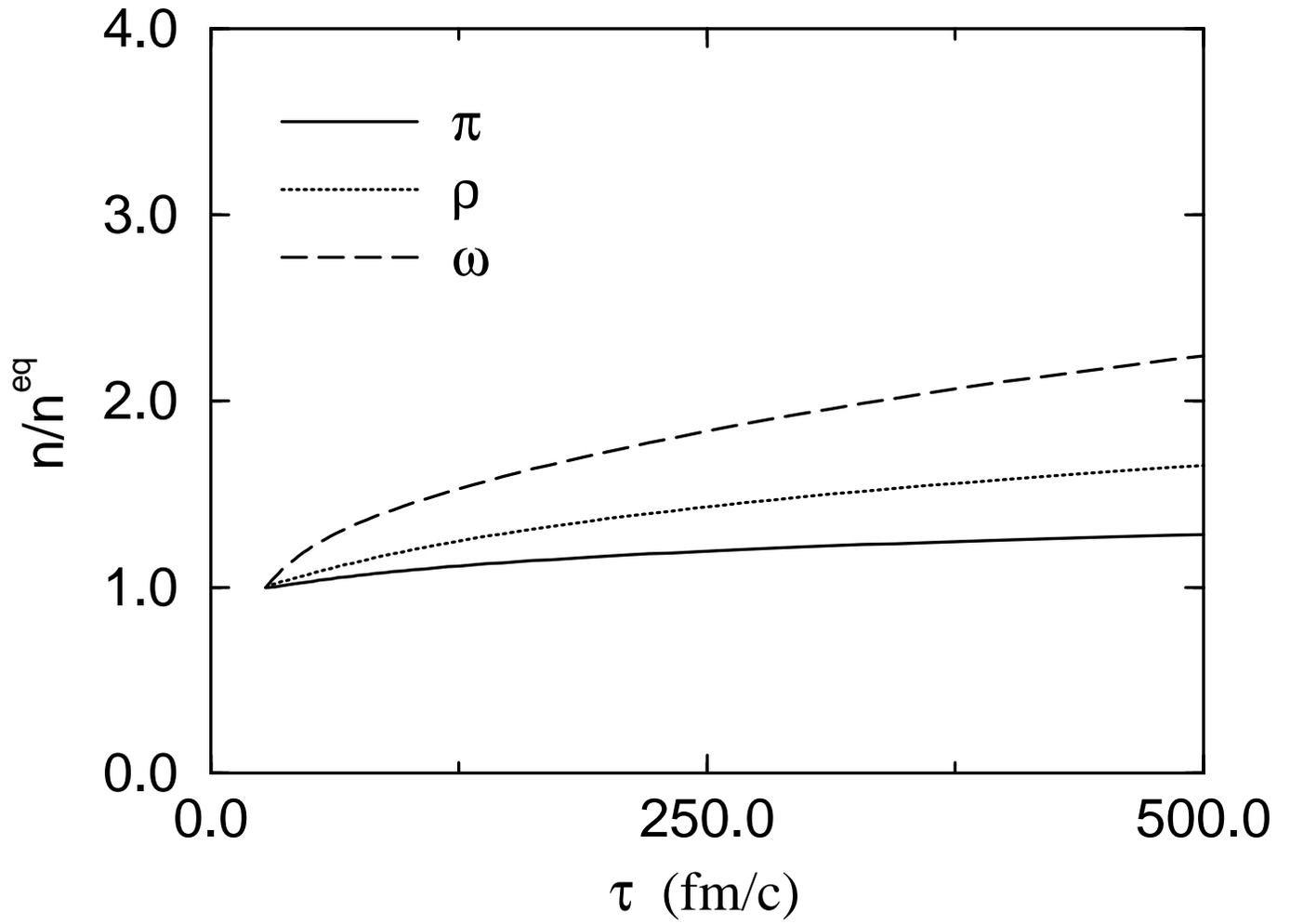
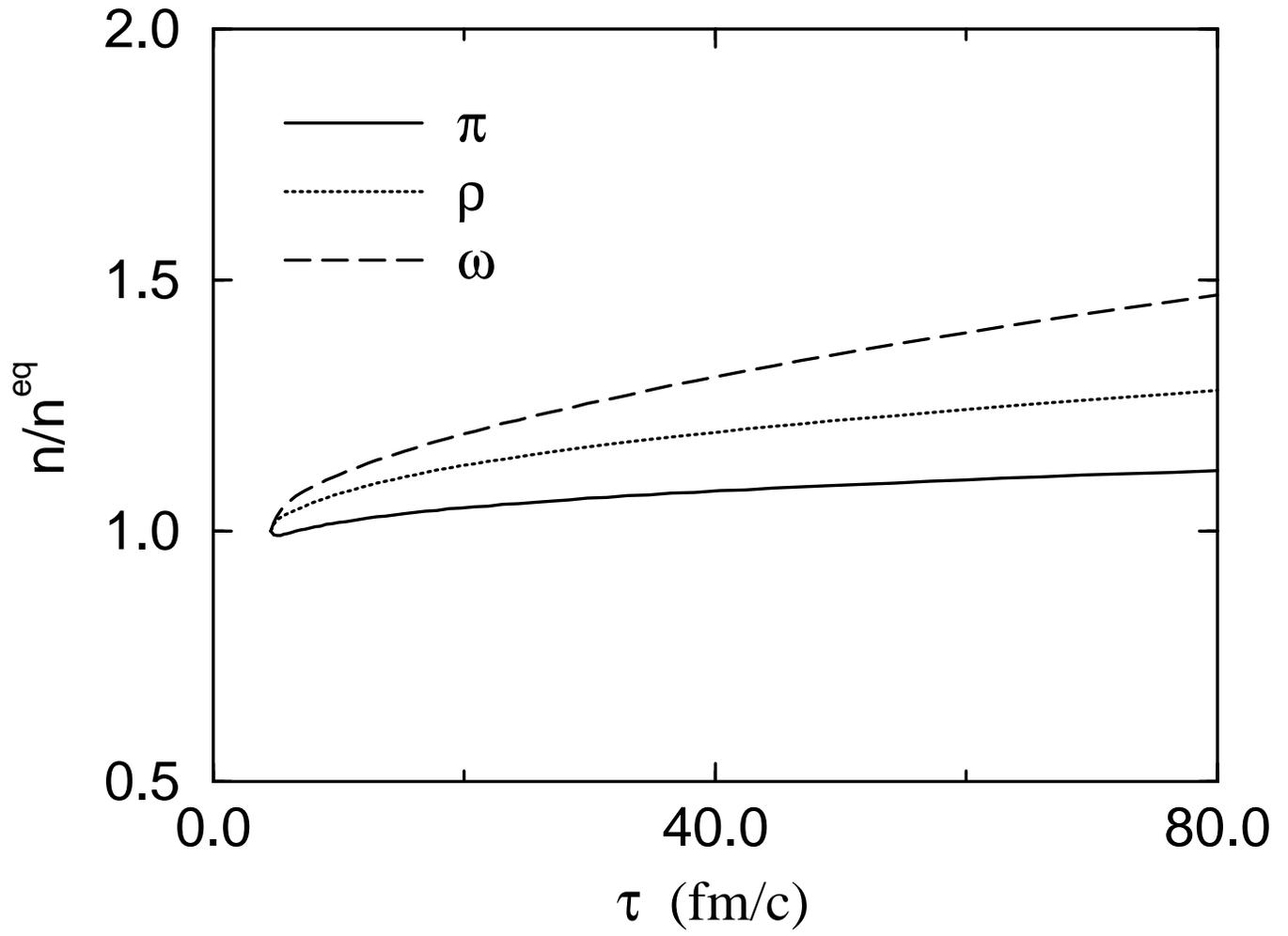
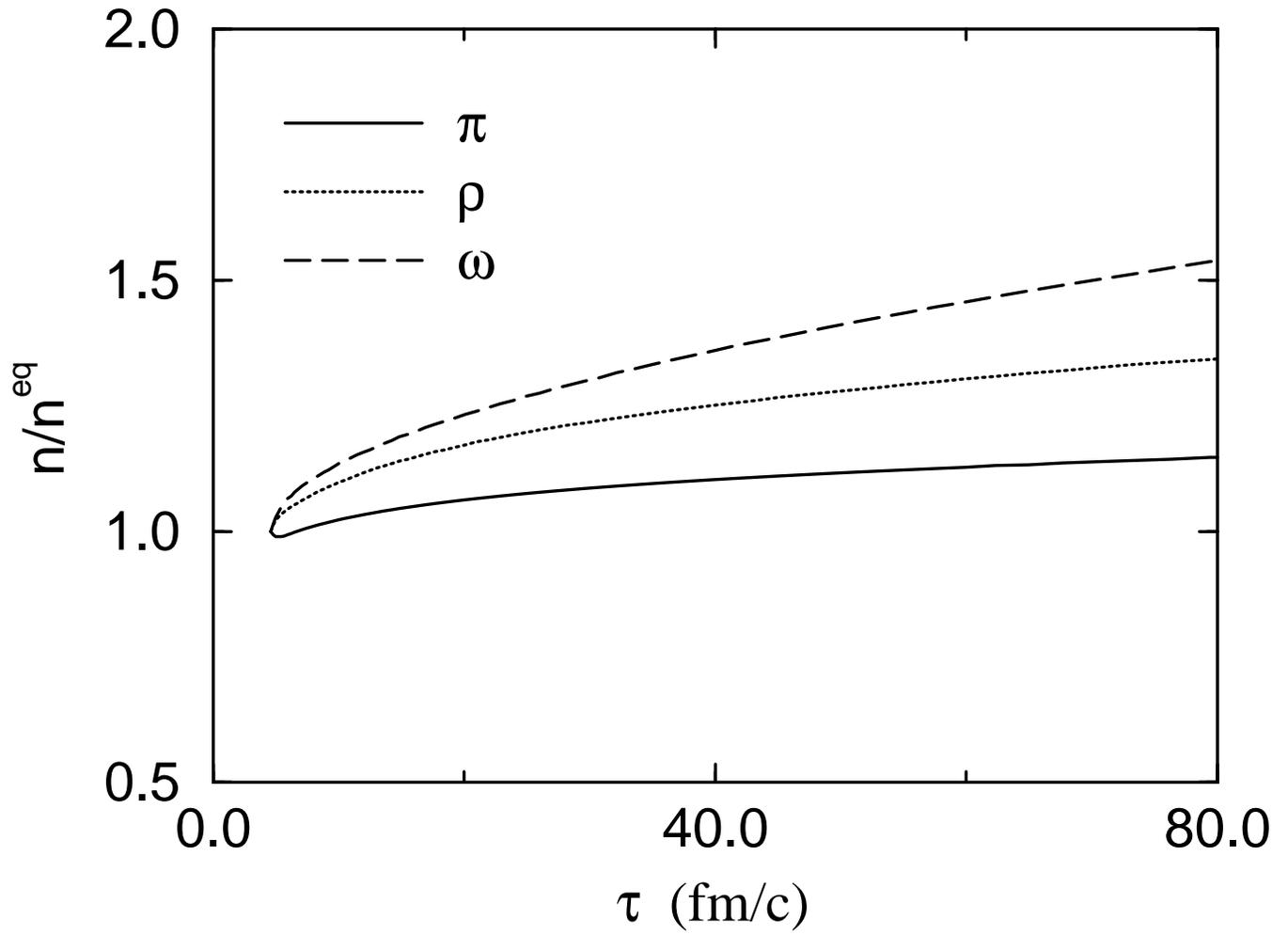
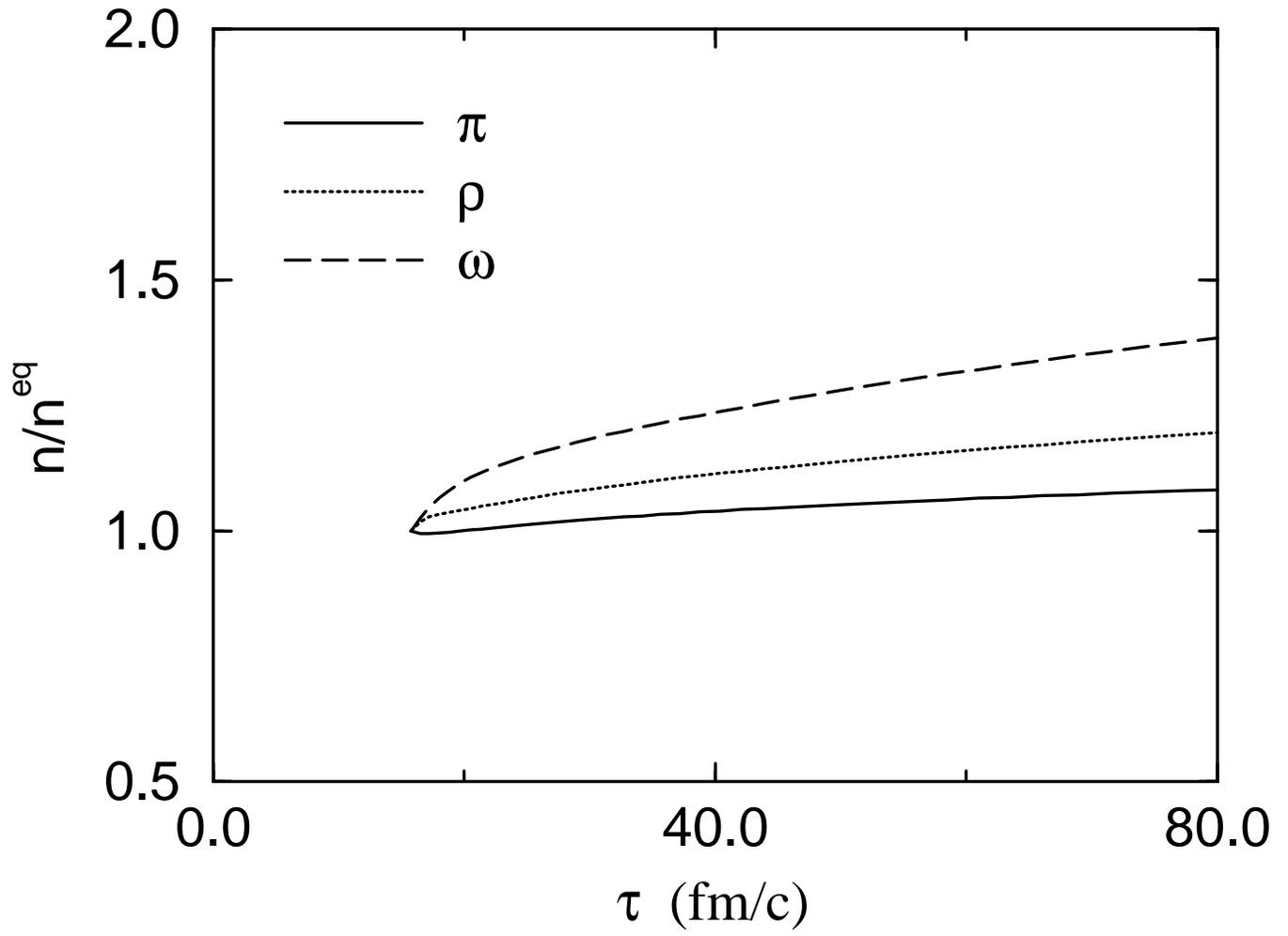
\end{document}